# AY Peg, a large amplitude Algol-type star

G. Boistel, S. Ferrand

Groupe Européen d'Observations Stellaires (GEOS)

**ABSTRACT**

The present paper gives new elements for the light variations of the EA variable star AY Peg, on the basis of new times of minimum performed visually and with ccd by members of GEOS between 1985 and 2018, and the ASAS-SN set of data available. On one hand, we can establish a new ephemeris with a possible quadratic term, and on the other hand, the amplitude of the primary minimum appears much deeper than the one given in GCVS. AY Peg varies between 13.1 and 15.6 magnitude at its primary eclipse.

## 1. INTRODUCTION

AY Peg is classified as an EA variable star in the GCVS (Samus et al., 2017) with the following elements:
- Coordinates J2000.0: 22h 00m 47.38s ; +34° 57' 48.1''
- Range 13.1-14.1 in V, spectrum A
- Ephemeris: Min I (hel. JD) = $2444462.565 + 2.43901 \times E$ (1)
- Duration of the primary eclipse: 20% of the period

The main references for the GCVS are two previous GEOS publications by Alain Figer (1978, 1980). In these papers, A. Figer established a first ephemeris for AY Peg and announced the algol-type for this eclipsing binary. Being quite faint, this star has remained rather understudied since these two GEOS studies.

We have re-explored the light variations of this star on the basis of new times of minimum extracted from our old individual visual estimates (performed with 256, 300 and 406mm diameter reflectors between 1985 and 2018) and new ccd determinations published mainly by BRNO, BBS and BAV observers (see table 1 for references). Anton Paschke has collected a list of known times of minimum in his O-C Gateway web site (http://var2.astro.cz/ocgate/index.php?lang=en).

Then, with the help of our old and new observations, we can establish a list of 39 times of minimum (table 1), and use data sets from the ASAS-SN Sky Patrol database (Shappee et al., 2014; Kochanek et al., 2017) to propose a renewed ephemeris (with a quadratic term). We can prove now that AY Peg is an EA, probably with a partial eclipse, without flat primary minimum, with a very faint secondary minimum (not yet observed), and with a primary amplitude larger than the one previously announced.

## 2. A NEW EPHEMERIS

From the 39 heliocentric times of minimum gathered for this paper (table 1), we can draw an O-C (in days) diagram (figure 1) computed without the photographic minimum from Meinunger (1980), and with the epoch based on the minimum observed by Stéphane Ferrand on July 30th, 2014. Uncertainties on visual estimates have been derived from light curve fits with cubic splines and Kwee and Van Woerden method, computed with the software PERANSO 3$^{©}$. Errors for the photographic minimum are simply a mean approximation between photoelectric and visual determinations; Meinunger (1980) does not give any uncertainty on his times of minimum in his papers published around 1980. This is due to the fact that these are not minimum measured with a significative number of data points but low brightness points measured on single photograph.

Within the main groups of observations on figure 1 we can identify the three « old » sets of minimum:

1°. The Meinunger's photographic ones;
2°. The GEOS EB1 group;
3°. The BRNO/BAV/BBSAG ccd minimum.

Between these second and third groups are the visual minimum observed by Alain Figer (FGR), Stéphane Ferrand (FND) and Guy Boistel (BTL) during the years 1985-1991, and more recently, the minimum observed by Stéphane Ferrand in 2014 and 2016 (visually), and Roland Boninsegna (BNN) in 2018 (with ccd device).

One can see that the O-Cs seem to be well fitted with a parabola as shown on figure 2 (left), while the residuals of the O-C computed with the quadratic ephemeris (2) are shown on figure 2 (right).

Between 1970s (E=-6000) and the years 2016-2018 (E close to +1000), the shift in the times of minimum is about +6,25 hours, as shown on figure 2 (left) with the ephemeris (1).

The residuals of the O-C illustrated by figure 2 (right) and computed with the ephemeris (2) are less than ± 0.05 days (± 1.2h) for recent minimum (E > -6 000), which is rather good, taking into account the short list of times of minimum we provide, and a primary minimum which is not so easy to obtain visually due to its low amplitude.

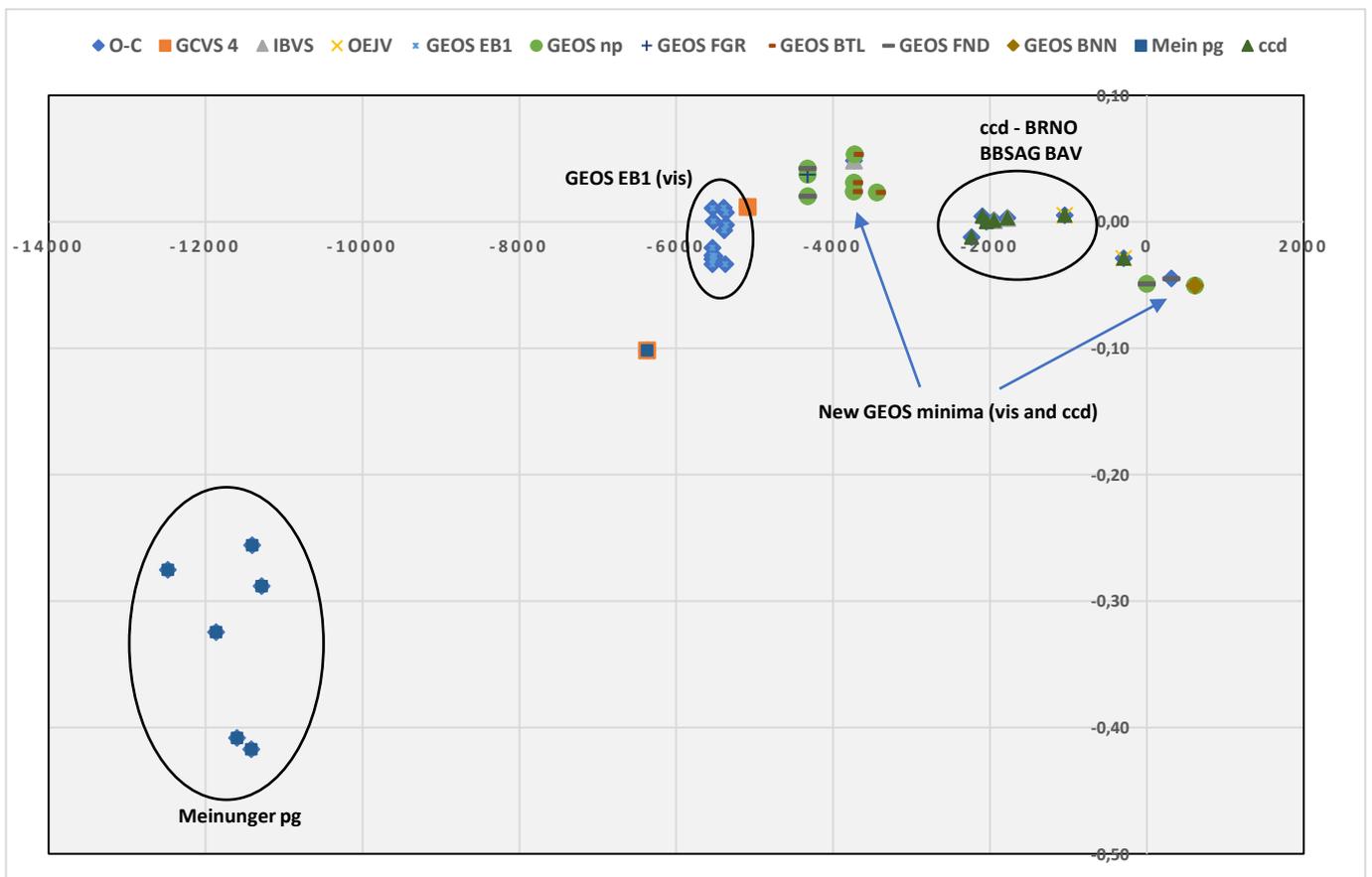

*Figure 1: O-C diagram as a function of cycle number E; units of O-C are days. We give the individual identification of each observer and each set of available observations.*

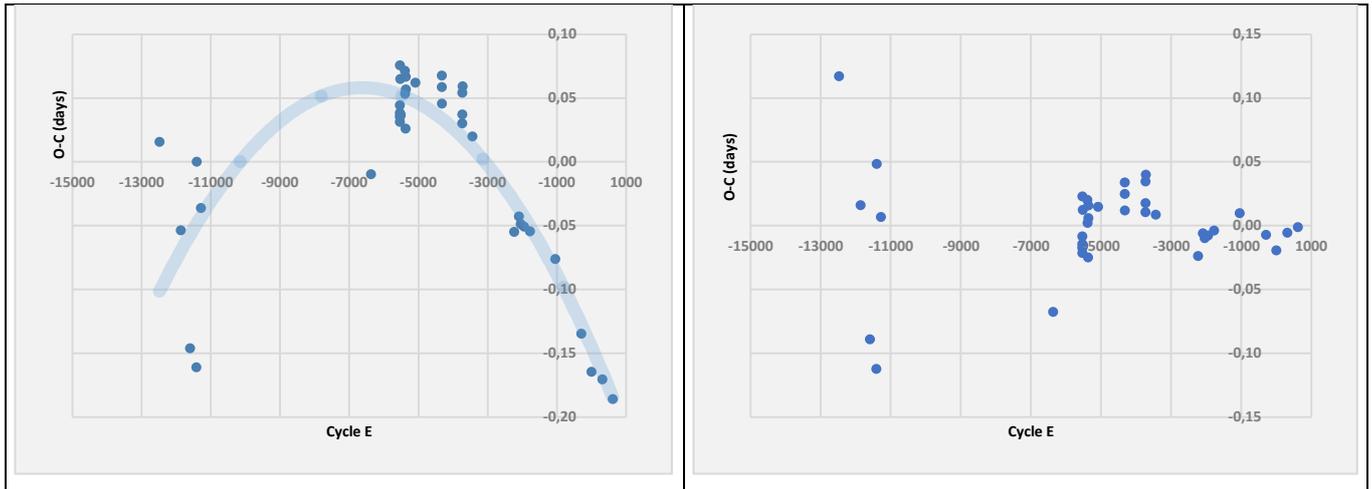

*Figure 2: left: parabolic fit on the O-C previously computed on ephemeris (1); Suspicion for a quadratic term in the ephemeris. Right: O-C residuals on the basis of the quadratic ephemeris (2).*

Then, we can adjust the O-C with a quadratic fit and establish a new ephemeris (2) with standard deviation:

**EPHEMERIS (2): MIN I (Hel. JD) = 2456869.520 + 2.438933 × E − 4.64.10$^{-9}$ × E²**
  ±.017  ±6  ±.42

To test this new ephemeris (2) we can compute phase diagrams for visual observations performed by Guy Boistel (BTL) in the years 1989-1991, with a 300mm-reflector, and Stéphane Ferrand (FND), with a 406 mm-reflector, during the years 2014-2016 (figure 3). The reduced amplitude of Boistel's light curve is easily explained by the fact that this primary amplitude is deeper than the previously expected magnitude of 14.2. A 300-mm reflector is not sufficient to observe visually that variable star at its primary minimum. Nevertheless, we can see that the ephemeris (2) fits these sets of visual estimates very well.

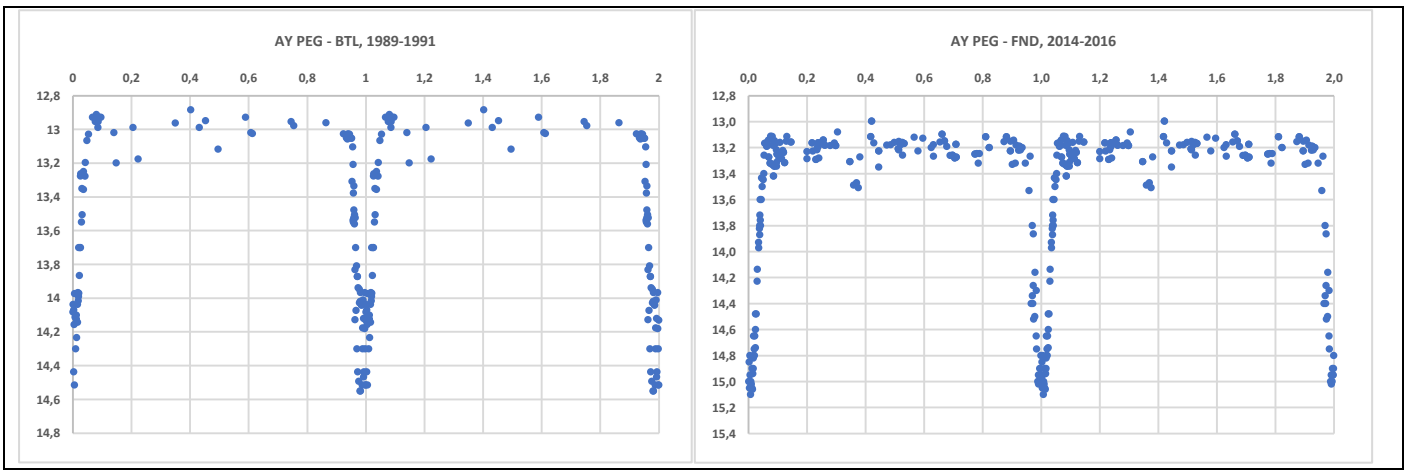

*Figure 3: Phase diagrams for BTL on left (1989-1991), and FND on right (2014-2016) computed on ephemeris (2). Left figure shows a deeper primary minimum.*

We can have a better proof of a deeper primary minimum with the phase diagram computed on the ASAS-SN Sky Patrol set of observations extracted from the AAVSO-VSX portal (link: https://asas-sn.osu.edu/variables/0346b007-7e82-5d09-8d05-b71b9a3b6886).
As shown on figure 4, our ephemeris fits the ASAS-SN set of observations as well. The primary amplitude is clearly shown larger than the expected amplitude, at about magnitude 15.7.

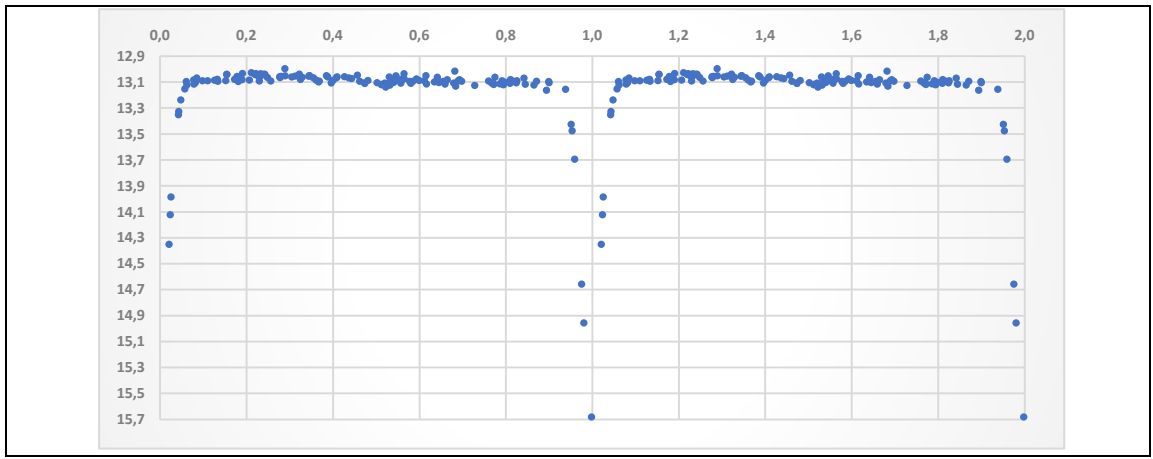

*Figure 4*: Phase diagram for ASAS Sky Patrol set of observations of AY Peg, computed on ephemeris (2).

### 3. A LARGER PRIMARY AMPLITUDE AND A PARTIAL ECLIPSE

The figures 3 (right) and 4 show that the amplitude of the primary minimum of AY Peg is deeper than the 1.0 magnitude amplitude (13.1-14.1) announced by the GCVS and Figer (1978). This primary amplitude is quite close to 2.5 magnitude, with a range from13.0/13.1 to 15.6 magnitude at least, taking into account ASAS-SN measurements of AY Peg. This result is in perfect agreement with Roger Diethelm's note in BBSAG Bulletin 123 (Blaettler et al., 2000): "*The amplitude determined from our unfiltered CCD data is considerably larger than the one given in the GCVS, namely 13.1 – 15.6 mag*".

We can state now that the eclipse is probably a partial one and there is no flat primary minimum (figures 3 and 4) until we have further very good determinations of the real primary minimum.

From FND's average phase diagram it is possible to deduce a new duration for the primary eclipse: $D \approx 0.137 \times P$ in agreement with available individual observed minima and the same determination from ASAS set of data (figure 4). Then, the duration of the primary eclipse is about equal to 14% of the period, a little smaller than the one announced in GCVS (20% of the period).

Figer (1980) gave a secondary minimum of 0.13 magnitude at phase 0.483, a clue for an eccentric orbit. However, the secondary minimum is still to be determined and its amplitude is, undoubtedly, smaller than 0.1 magnitude. Figure 4 (ASAS-SN measurements) seems to show a rather shallow secondary minimum at phase 0.5, but the uncertainty remains too large to conclude.

### 4. A NEW VISUAL FINDING CHART FOR AY PEG

We give here a finding chart with a new sequence for visual observers (figure 5), from the online AAVSO plotter tool and GAIA catalogs of stars (for the magnitudes). AY Peg is visually easy to find and to observe; it is surrounded by a typical trapezoid of bright stars. Due to the large amplitude, we propose a new finding chart with a comparison star close to the 16$^{th}$ magnitude. Therefore, a 400-mm diameter reflector is highly recommended for visual observation of this star.

Luckily, there is no known variable star in a field of 20' radius around AY Peg and any star can be chosen to perform Argelander-type comparisons. We propose the chart below with 5 comparison stars. The star labeled L was suspected to be variable by Alain Figer in the years 1970-1980; that needs to be verified. If it is not variable, the star L is very suitable for visual comparisons.

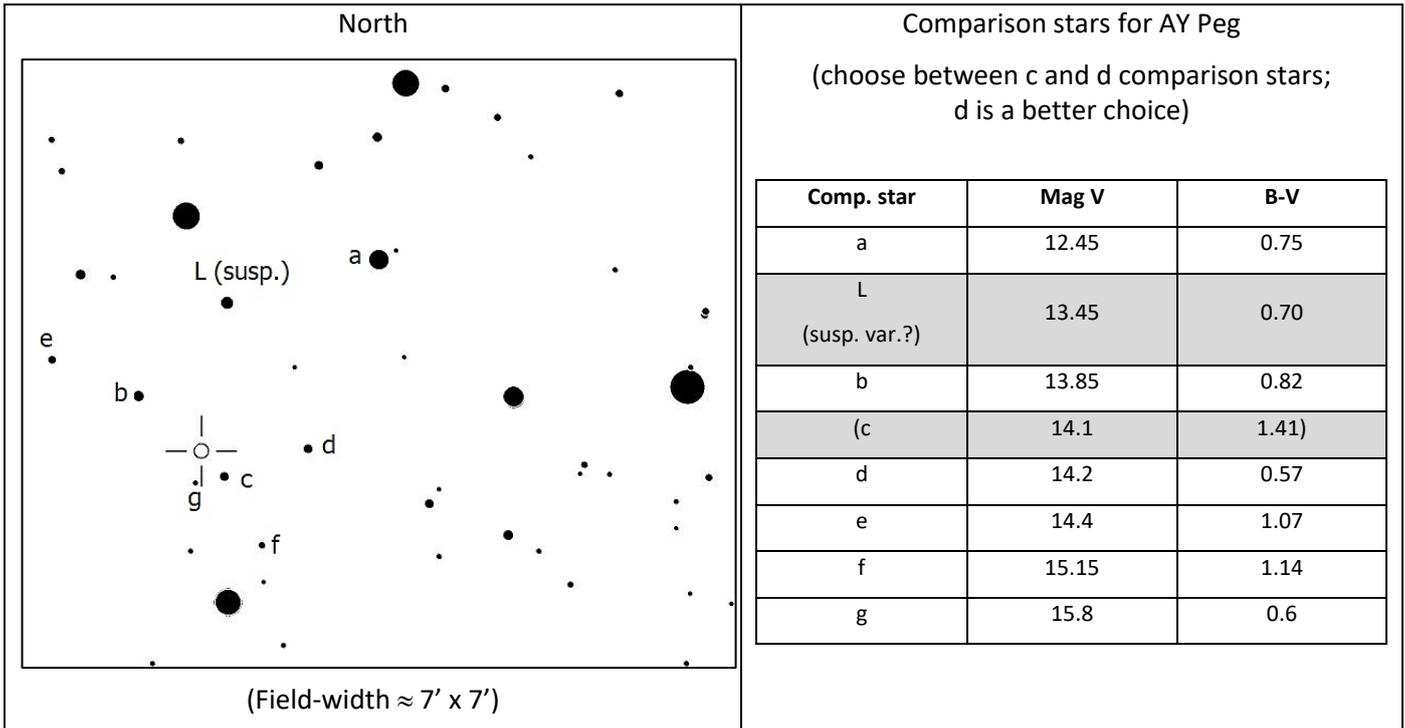

*Figure 5: Chart and comparison stars for AY Peg.*

### 5. ACKNOWLEDGEMENTS

Table 1: List of the 39 observed minima.

| Hel.JD-2400000 | Error (days) | Method | Observer | Reference |
|---|---|---|---|---|
| 26433.4750 | 0.0030 | pg |  | Meinunger, 1980 |
| 27933.3870 | 0.0030 | pg |  | Meinunger, 1980 |
| 28584.5060 | 0.0030 | pg |  | Meinunger, 1980 |
| 29028.3880 | 0.0030 | pg |  | Meinunger, 1980 |
| 29055.3780 | 0.0030 | pg |  | Meinunger, 1980 |
| 29350.4600 | 0.0030 | pg |  | Meinunger, 1980 |
| 41333.2640 | 0.0030 | pg |  | Meinunger, 1980 |
| 43367.4260 | 0.0030 | vis | Figer A. | Figer, 1978 |
| 43367.4300 | 0.0030 | vis | Mailler R. | Figer, 1978 |
| 43367.4330 | 0.0030 | vis | Figer A. | Figer, 1978 |
| 43367.4390 | 0.0030 | vis | Ralincourt P. | Figer, 1978 |
| 43372.3483 | 0.0030 | vis | Figer A. | Figer, 1978 |
| 43396.7275 | 0.0030 | vis | Figer A. | Figer, 1978 |
| 43401.5767 | 0.0030 | vis | Figer A. | Figer, 1978 |
| 43428.4065 | 0.0030 | vis | Figer A. | Figer, 1978 |
| 43718.6812 | 0.0030 | vis | Figer A. | Figer, 1978 |
| 43733.2970 | 0.0030 | vis | Figer A. | Figer, 1978 |
| 43767.4157 | 0.0030 | vis | Figer A. | Figer, 1978 |
| 43784.5293 | 0.0030 | vis | Figer A. | Figer, 1978 |
| 43789.3974 | 0.0030 | vis | Figer A. | Figer, 1978 |
| 44462.5650 | 0.0030 | pg | GCVS | GCVS |
| 46323.5140 | 0.0050 | vis | Figer A. | Present paper |
| 46323.5182 | 0.0032 | vis | Ferrand S. | Present paper |
| 46328.3790 | 0.0030 | vis | Ferrand S. | Present paper |
| 47762.4906 | 0.0098 | vis | Boistel G. | Present paper |
| 47767.3770 | 0.0030 | vis | Boistel G. | Present paper |
| 47767.3940 | 0.0030 | vis | Martignoni M. | Diethelm, 2003 |
| 47789.3500 | 0.0030 | vis | Boistel G. | Present paper |
| 48479.5460 | 0.0030 | vis | Boistel G. | Present paper |
| 51423.3370 | 0.0010 | ccd | Paschke A. | O-C Gateway |
| 51757.4913 | 0.0010 | ccd | Diethelm R. | Blaettler et al., 2000 |
| 51884.3130 | 0.0010 | ccd | Diethelm R. | Blaettler et al., 2001 |
| 52118.4544 | 0.0010 | ccd | Agerer F | Agerer, Hubscher, 2002 |
| 52535.5187 | 0.0010 | ccd | Diethelm R. | Diethelm, 2003 |
| 54318.4016 | 0.0001 | ccd | Lehky M. | Prat et al., 2009 |
| 56152.4666 | 0.0010 | ccd | Trnka J. | Hoňková et al., 2013 |
| 56869.5010 | 0.0030 | vis | Ferrand S. | Present paper |
| 57635.3392 | 0.0073 | vis | Ferrand S. | Present paper |
| 58369.4610 | 0.0060 | pe | Boninsegna R. | Present paper |